\documentclass{appolb}
\usepackage[utf8]{inputenc}
\usepackage{epsf}
\usepackage{latexsym,amssymb,euscript}
\usepackage[dvips]{graphicx}
\usepackage[numbers,sort&compress]{natbib}
\usepackage{amsmath}
\usepackage{nicefrac}
\usepackage{slashed}
\usepackage{booktabs}
\usepackage{hyperref}
\usepackage{braket}
\usepackage{chngcntr}
\usepackage{bbold}
\usepackage{graphics}
\usepackage{graphicx}
\usepackage{mciteplus}
\usepackage{caption}
\usepackage{subcaption}
\usepackage{pdfpages}
\usepackage[titletoc]{appendix}
\graphicspath{{./figures/}}
\hypersetup{
 linktocpage = true,
 urlcolor = black,
 colorlinks = true,
 linkcolor = urlblue,
 anchorcolor = urlblue,
 citecolor = urlblue,
 pdfstartview = {XYZ null null 1.25} 
           }
\usepackage{pstricks}
\usepackage{color}
\usepackage{xcolor}
\definecolor{urlblue}{rgb}{0.2,0.4,0.7}
\definecolor{citegreen}{rgb}{0,0.4,0.2}
\definecolor{linkred}{rgb}{0.9,0.2,0.1}
\usepackage{float}
\definecolor{orcidlogocol}{HTML}{A6CE39}
\usepackage{fancyhdr}

\newcommand{\tarr}{
\begin{array}}
\newcommand{\earr}{\end{array}}

\begin{document}
 \eqsec  
\title{Next-to-Soft Virtual Resummation for QCD Observables
\thanks{Presented by A. H. Ajjath at ``Diffraction and Low-$x$ 2022'', Corigliano Calabro (Italy), September 24-30, 2022.}%
}
\author{A. H. Ajjath$^{1}$, Pooja Mukherjee$ ^{2} $, V. Ravindran$^{3}$, Aparna Sankar $ ^{4} $ and Surabhi Tiwari$ ^{5} $
\address{\vspace{0.4cm}$^1$Laboratoire de Physique Th\'eorique et Hautes Energies (LPTHE), UMR 7589, Sorbonne Universit\'e et CNRS, 4 place Jussieu, 75252 Paris Cedex 05, France\\
	$^2$Bethe Center for Theoretical Physics, Universit\"at Bonn, 53115 Bonn, Germany\\ 
 $^3$The Institute of Mathematical Sciences, HBNI, IV Cross Road, Taramani, Chennai 600113, India\\
	$^4$Physik-Department, Technische Universität München, James-Franck-Strasse 1, 85748 Garching,
Germany,Max-Planck-Institut für Physik, Föhringer Ring 6, 80805 München, Germany\\  
	$^5$Institut f\"{u}r Theoretische Teilchenphysik, Karlsruhe
Institute of Technology (KIT),
Wolfgang-Gaede Stra{\ss}e 1, 76128 Karlsruhe, Germany}
}
\maketitle
\begin{abstract}
We present a framework for resumming the contributions from soft-virtual and next-to-soft virtual (NSV) logarithms. Numerical impact for these resummed predictions are discussed for the inclusive cross section for Drell-Yan di-lepton process up to  next-to-next-to leading logarithmic accuracy, restricting to only diagonal partonic channels. 
\end{abstract}
\section{Introduction}    
Precise predictions of QCD observable not only shed light on the new physics signatures, but they also reveal the rich mathematical structures in the underlying gauge theories. 
Performing higher order predictions in perturbative QCD involves complex Feynman loop integrals and many body phase space integrals. Due to the complexity in the computations, it is a general practise to look for alternative approaches by taking certain approximations. One good alternative is expanding the perturbative series around the production threshold, defined in terms of partonic scaling variable $z = \frac{Q^2}{\hat s} \approx 1$, where $Q$ denotes the invariant mass of the final state system produced in the partonic reactions with their centre of mass energy $\hat s$. Such an expansion also helps to understand the logarithmic structure in higher order perturbative results. The leading term in the expansion, often called soft-virtual (SV) corrections, involves contributions from soft gluon emissions along with the Feynman loop corrections. At the production threshold, these soft gluon emissions results in large logarithms of the form  $\big(\frac{\ln^j(1-z)}{1-z}\big)_+,j\ge0$, which needs to be resummed in order to get reliable predictions. The resummation framework for the SV logarithms are well-known \cite{Sterman:1986aj,
Catani:1989ne,
Moch:2005ba,Bonvini:2012sh,Catani:2014uta,Ajjath:2020rci}
to the third order in logarithmic accuracy, thanks to the numerous efforts along this direction. 
 
Recently, there has been many efforts to study the structure of next-to-leading term in the threshold expansion, which are of the form $\ln^j(1-z),j\ge0$ and their resummation to all order in perturbation theory (See
\cite{Laenen:2008ux,Grunberg:2009yi,Moch:2009hr,Bonocore:2014wua,Beneke:2019oqx,DelDuca:2017twk,deFlorian:2014vta,Das:2020adl}). These  contributions are called next-to-SV (NSV) logarithms. In \cite{Ajjath:2020ulr,Ajjath:2022kyb}, we propose a framework for the first time, to study the resummation of NSV logarithms beyond leading logarithmic (LL) accuracy for the color singlet productions, restricting to only the diagonal partonic channels. In the present article, we report the numerical impacts of the NSV logarithms to third order in logarithmic accuracy for the case of Drell-Yan di-lepton process at LHC.

\section{Next-to-Soft Virtual Framework}\label{sec:DY}
In QCD improved parton model, the differential invariant mass distributions for a heavy colorless final states produced in hadron collisions takes form of a convoluted integral:
\begin{eqnarray}
{d \sigma \over dQ}(q^2,\tau) = \sigma^{(0)} \int_{\tau}^1 {d z \over z } \tilde \Phi_{ab}\left({\tau \over z},\mu_F^2 \right) \Delta_{ab}(q^2,\mu_F^2.z) \,.
\end{eqnarray}
where $\sigma^{(0)}$ is the born-cross section. The partonic flux $\tilde \Phi_{ab}$ is defined to be 
\begin{eqnarray}
\tilde \Phi_{ab}\left({\tau \over z},\mu_F^2\right) = \int_{\tau \over z}^1 {d y \over y } f_a(y,\mu_F^2) f_b\left({\tau \over z y},
\mu_F^2\right) 
\end{eqnarray}
with the factorisation scale $\mu_F$ and the incoming parton distribution function $f_c$. Also, $\tau=q^2/S$ is the hadronic scaling variable with hadronic centre of mass energy $S$ and $a,b=q,\overline q,g$ refer to incoming partonic states. The $\Delta_{ab}$ is the perturbatively calculable coefficient
functions,
which can be decomposed
accoding to their singular behaviour:
\begin{align}\label{deltadecompose}
\Delta_{ab}(q^2,\mu_F^2,z) =  \delta_{a b} \Delta_{a\overline a}^{SV}(q^2,\mu_F^2,z) + \Delta_{ab}^{NSV}(q^2,\mu_F^2,z) +  \Delta_{ab}^{reg}(q^2,\mu_F^2,z) \,.
\end{align}
Each of these terms is perturbatively expanded in terms of renormalised strong coupling constant $a_s = g_s^2/16 \pi^2$. For $J=SV, NSV, reg$ we have 
  $ 
      \Delta_{ab}^{(J)}(q^2,\mu_F^2,z) = \sum_{i=0}^\infty a_s^i(\mu_R^2) \Delta_{ab}^{J,(i)}(q^2,\mu_R^2,\mu_F^2,z).
 $ 
 where $\mu_R$ refers to renormalisation scale.
The first term in \eqref{deltadecompose} is the SV corrections, which gets contributions from following distributions:
\begin{align}\label{SV}
 \Delta^{SV,(i)}_{ab}(z) = \delta_{ab}\Big( \Delta_{a \overline a,\delta}~ \delta(1-z)
+ \sum_{k=0}^{2i-1}  \Delta^{(i)}_{ a \overline a,{\cal D}_k} {\cal D}_k(z) \Big)\,.
\end{align}
The second term comprises of $\ln^k(1-z)$
$,k=0,1,\cdots,\infty$,
\begin{align}
\label{DeltaR} 
\Delta_{ab}^{NSV,(i)}(z) = \sum_{k=0}^{2 i-1} \Delta_{ab,k}^{reg,(i)} \ln^k(1-z) \,.
\end{align}
and the rest are regular, of the form $(1-z)^m,m=1,\cdots,\infty$, when $z$ approaches 1.

The $z$-space coefficients in the above SV and NSV contributions involve convolutions, which are convenient to perform in Mellin $N$-space. The limit $z\rightarrow 1$, then translates to large $N$. These large logarithms with $a_s$ produce ${\mathcal O}(1)$ terms at every order in $a_s$ spoiling the truncation of perturbative series. Performing resummation resolves this by reorganising the series in terms of $\omega = 2 a_s(\mu_R^2) \beta_0 \ln N$ at every order. The well-known formula for the SV resummation reads \cite{Sterman:1986aj,Catani:1989ne}:
\begin{eqnarray}
\label{resumgen}
\lim_{N\rightarrow \infty} \ln \Delta_{c\overline c,N}^{SV} =
 \ln \tilde g^c_0(a_s(\mu_R^2))+
\ln N g^c_1(\omega) + \sum_{i=0}^\infty a_s^i(\mu_R^2) g^c_{i+2}(\omega) \,.
\end{eqnarray}
where $\Delta_{c \overline c,N} = \int_0^1 dz z^{N-1} \Delta_{c \overline c}(z)$.
In \eqref{resumgen} the resum coefficients $g^{c}_i(\omega)$ are universal and $\tilde g^{c}_0(a_s(\mu_R^2))$ collects $N$ independent terms. Inclusion of successive terms in the expansion predicts the leading-logarithms (LL),
next-to-LL (NLL), next-to-NLL (NNLL) and so on to all orders in $a_s$. Including these higher logarithmic corrections improve the fixed order results.   

Following the formalism described in \cite{Ajjath:2020ulr,Ajjath:2022kyb,Ajjath:2021lvg,Ajjath:2021bbm}
we systematically resum NSV logarithms for the inclusive Drell-Yan di-lepton process, restricting to only the diagonal channels. In addition to threshold $\log N$, we include the ${\cal O}(1/N)$ terms in large $N$ limit, in order to resum SV and NSV logarithms. Similar to SV case in \eqref{resumgen}, the NSV resum formula reads as
\begin{align}
\label{PsiNSVN}
 \lim_{N\rightarrow \infty} \ln \Delta_{c\bar c,N}^{NSV} = {1 \over N} 
\sum_{i=0}^\infty a_s^i(\mu_R^2) \Big ( \bar g_{i+1}^c(\omega)
+ \sum_{k=0}^{i} h^c_{ik}(\omega)~ \ln^k N.\Big)\,,
\end{align}
with NSV resum coefficients $\bar g^q_i(\omega)$ and $h^q_{ik}(\omega)$. These coefficients for the Drell-Yan process to NNLL are presented in the appendices of \cite{Ajjath:2021lvg}. In order to avoid double counting threshold logarithms, we finally match the resummed results in the $N$-space to the fixed order corrections  
\begin{align}
    \sigma_N^{\rm {N^nLO+\overline {\rm N^nLL}}} &= 
\sigma_N^{\rm {N^nLO}} +
\sigma^{(0)} 
\sum_{ab}
  \int_{c-i\infty}^{c+i\infty} \frac{dN}{2\pi i} (\tau)^{-N} \delta_{a\overline b}f_{a,N}(\mu_F^2) f_{b,N}(\mu_F^2) \nonumber\\
&\times \Big( \Delta_{c\bar c,N} \bigg|_{\overline {\rm {N^nLL}}} - {\Delta_{c\bar c,N}}\bigg|_{tr ~\rm {N^nLO}}     \Big)  \,.
\end{align}\label{eq:matched}
where $\sigma_N^{\rm N^nLO}$ is the Mellin moment of $d\sigma/dQ$ to the $n^{th}$ order in $a_s$. Also, $\overline{\rm  N^nLL}$ denotes the SV+NSV resummation, while N$^n$LL refers to the resummation of only SV logarithms. 
\section{Phenomenology}\label{sec:pheno}
The numerical setup we use for our study are detailed in \cite{Ajjath:2021lvg}. In brief, we choose the centre of mass energy 13 TeV at LHC with $\tt{MMHT2014}$ parton densities, the $a_s$ is evolved to $\mu_R$ in $\overline{\text{MS}}$-scheme and the electro-weak parameters are chose to be: Z-boson mass = 91.1876 \text{GeV} and width=2.4952\text{GeV},
$\text{sin}^2\theta_w=0.22343$ and fine structure constant  $ \alpha=1/128$.
\begin{figure}[ht]
\includegraphics[width=0.45\textwidth]{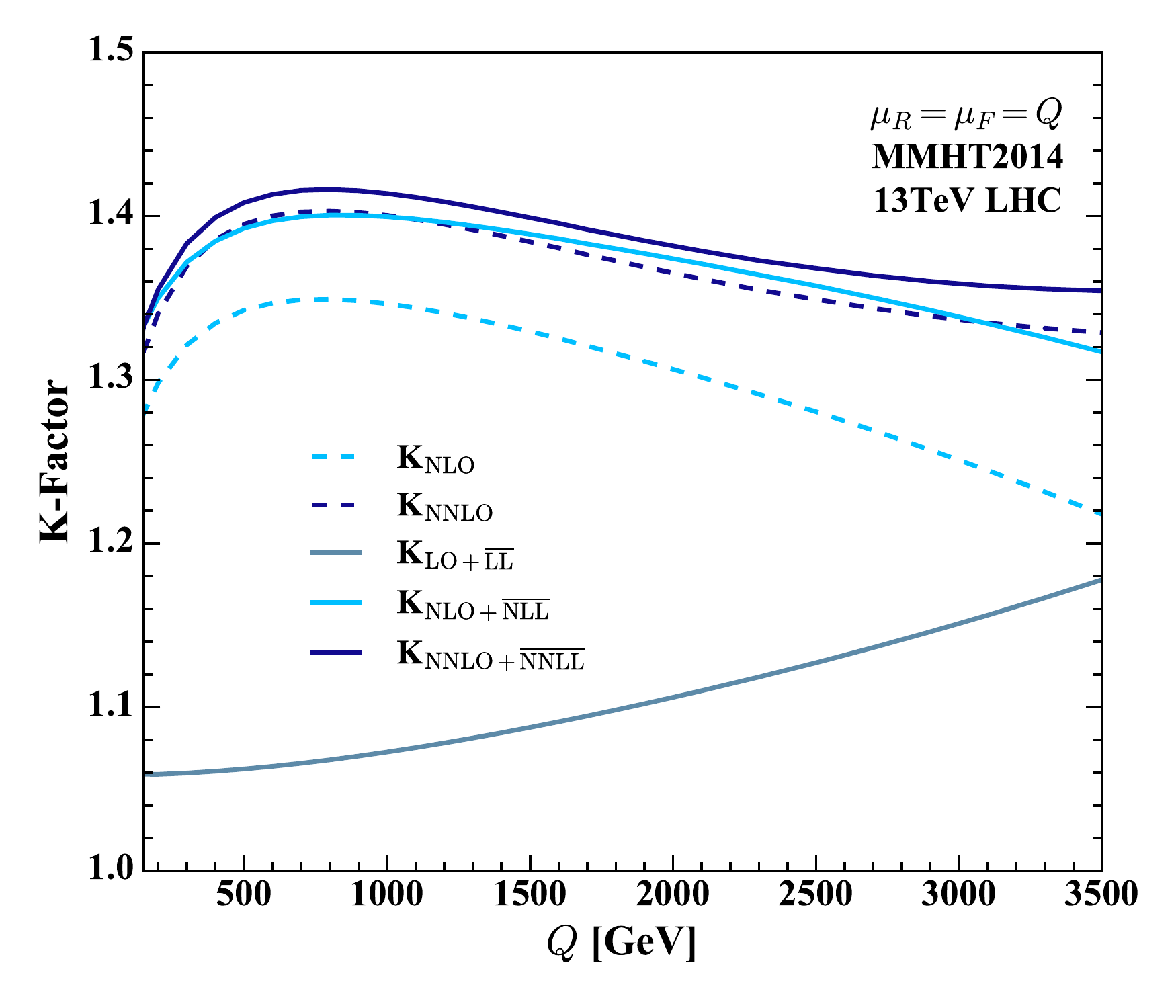}
\includegraphics[width=0.45\textwidth]{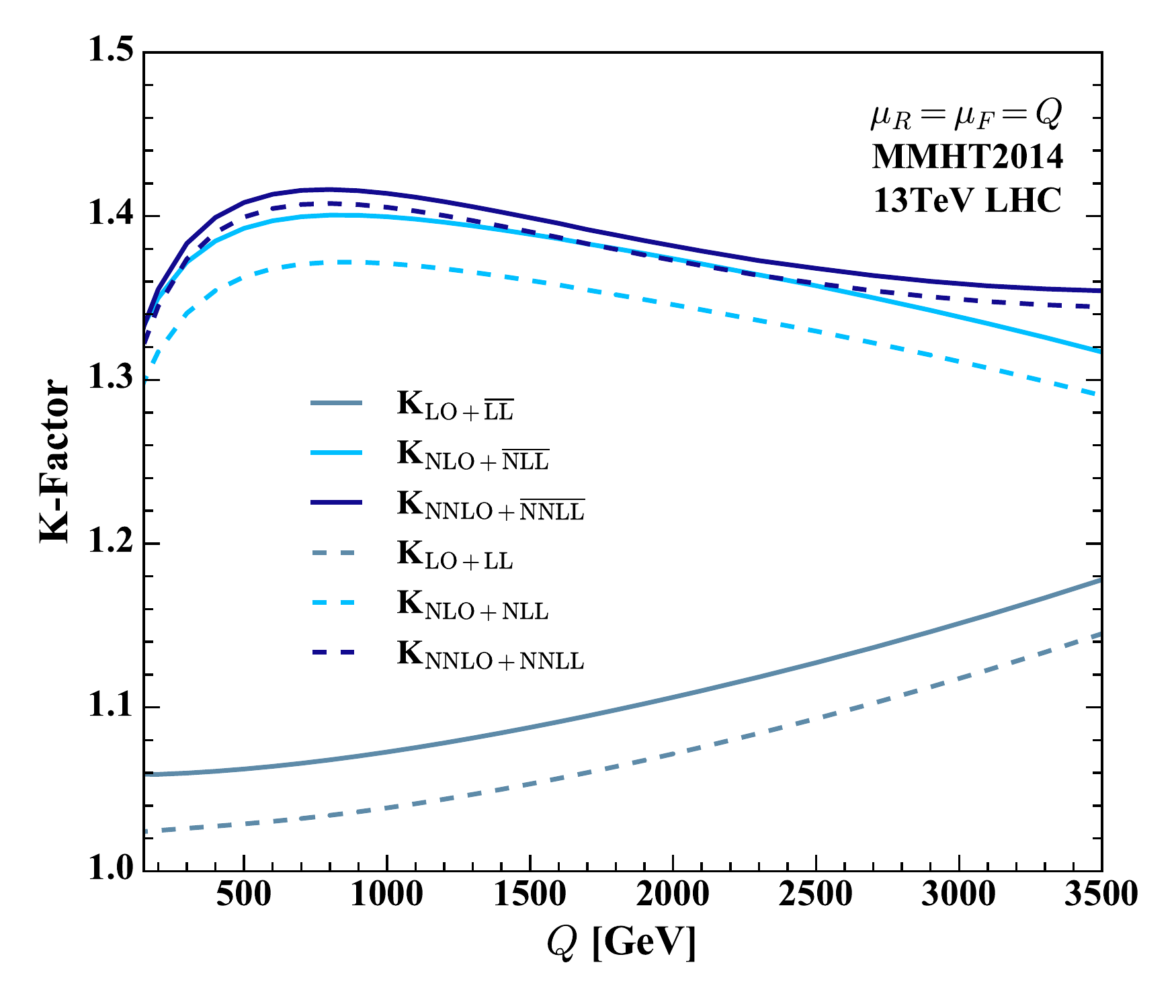}
\caption{\small{K-factors (left) till NNLO+$\overline{\rm{NNLL}}$ level at the central scale $Q=\mu_R=\mu_F$ and (right) for SV and NSV comparison}}
\label{KallN}
\end{figure}

We begin with comparing fixed order corrections to the NSV resummed predictions, using the  $``$K-factor" defined by
$ 
 \mathrm{K} \left(Q\right) = {\dfrac{d\sigma}{dQ}}/{\dfrac{d\sigma^{\text{LO}}}{dQ}} 
$ at $\mu_R=\mu_F=Q$.
 It is clear from Fig.\ref{KallN} (left panel) that the resummed predictions improve the fixed order results.  
Quantitatively, for example at Q=2000 GeV, the inclusion of ${ \overline {\rm NLL}}$ enhance the NLO by $5.2\%$ and ${ \overline {\rm NNLL}}$ modifies NNLO by $1.2\%$. Further, ${\rm NLO + \overline{NLL}}$ curve is closer to ${\rm NNLO}$, indicating that the inclusion of higher logarithms mimics the entire second order contributions.
To further see the effects of NSV logarithms in particular, we compare them against SV resummed results in right panel of Fig.\ref{KallN}. In higher orders, both SV and NSV resum results are found to be closer, accounting to the better perturbative convergence upon including NSV effects.

\begin{figure}[ht]
\centering
\includegraphics[width=0.47\textwidth]{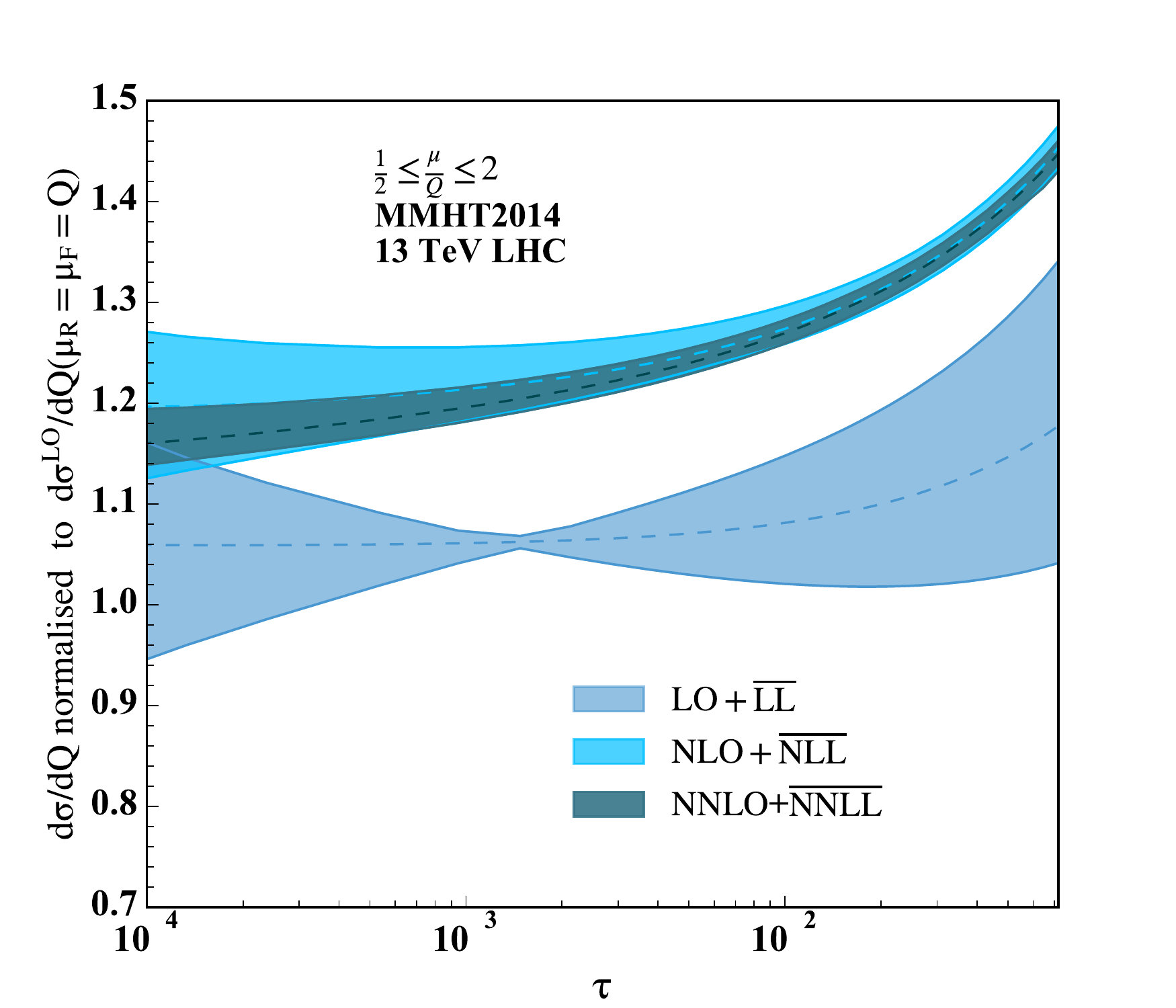}
\caption{7-point scale variation of the resummed results for the central scale choice $(\mu_R,\mu_F) = (1,1)Q$ for $13$ TeV LHC.}
\label{7ptNall}
\end{figure}

To assess the impact of renormalisation and factorisation scales, we estimate the error using canonical 7-point variation, with  $\frac{1}{2} \leq \big(\frac{\mu_R}{Q}, \frac{\mu_F}{Q}\big) \leq 2$, excluding the extreme points (0.5,2) and (2,0.5). This is depicted in Fig.\ref{7ptNall}, where the resummed results are found to be not much improved. The reason could be due to the lack of off-diagonal counter part, which will be evident in subsequent analysis.

\begin{figure}[!ht]
\hspace{0.5cm}
\includegraphics[width=0.45\textwidth]{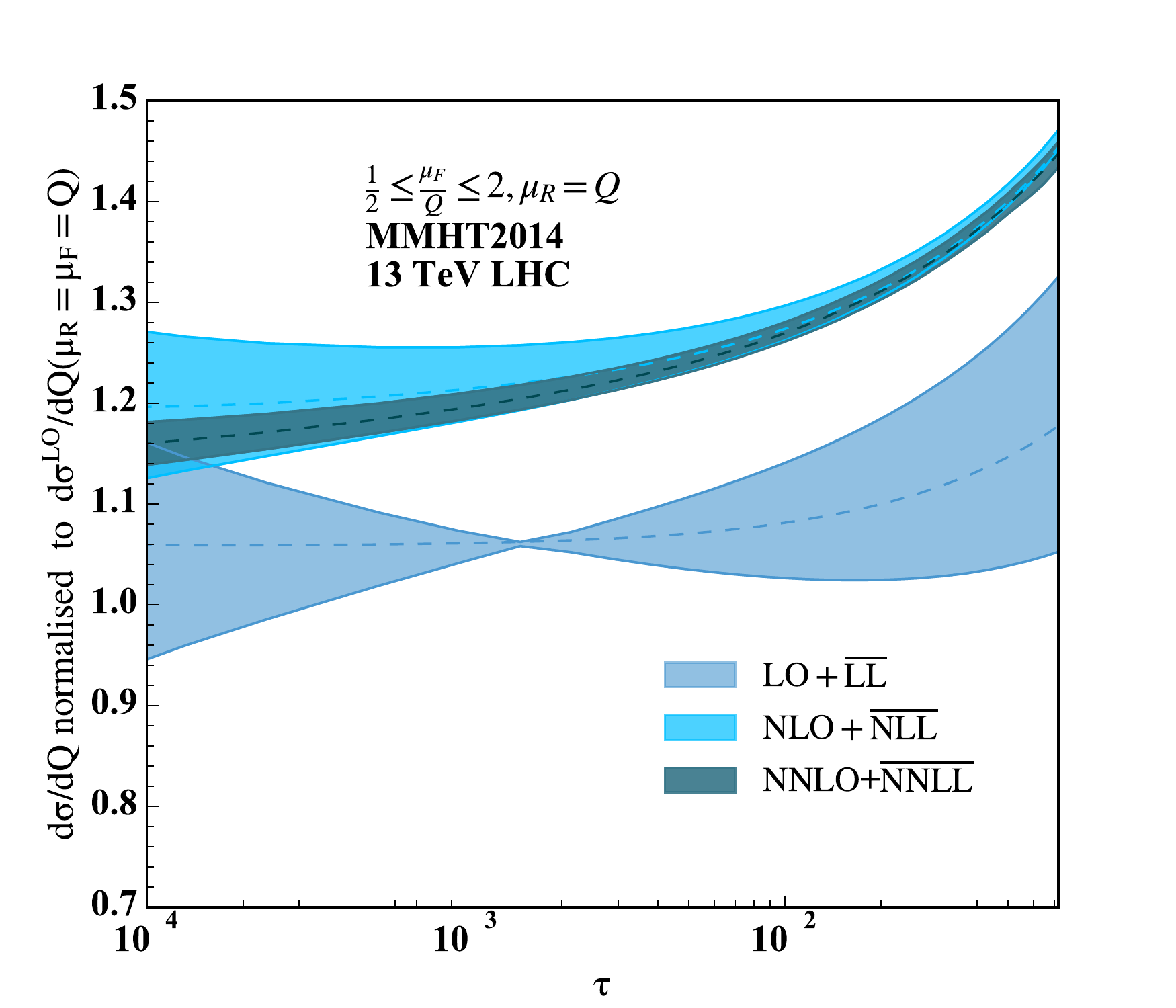}
\includegraphics[width=0.45\textwidth]{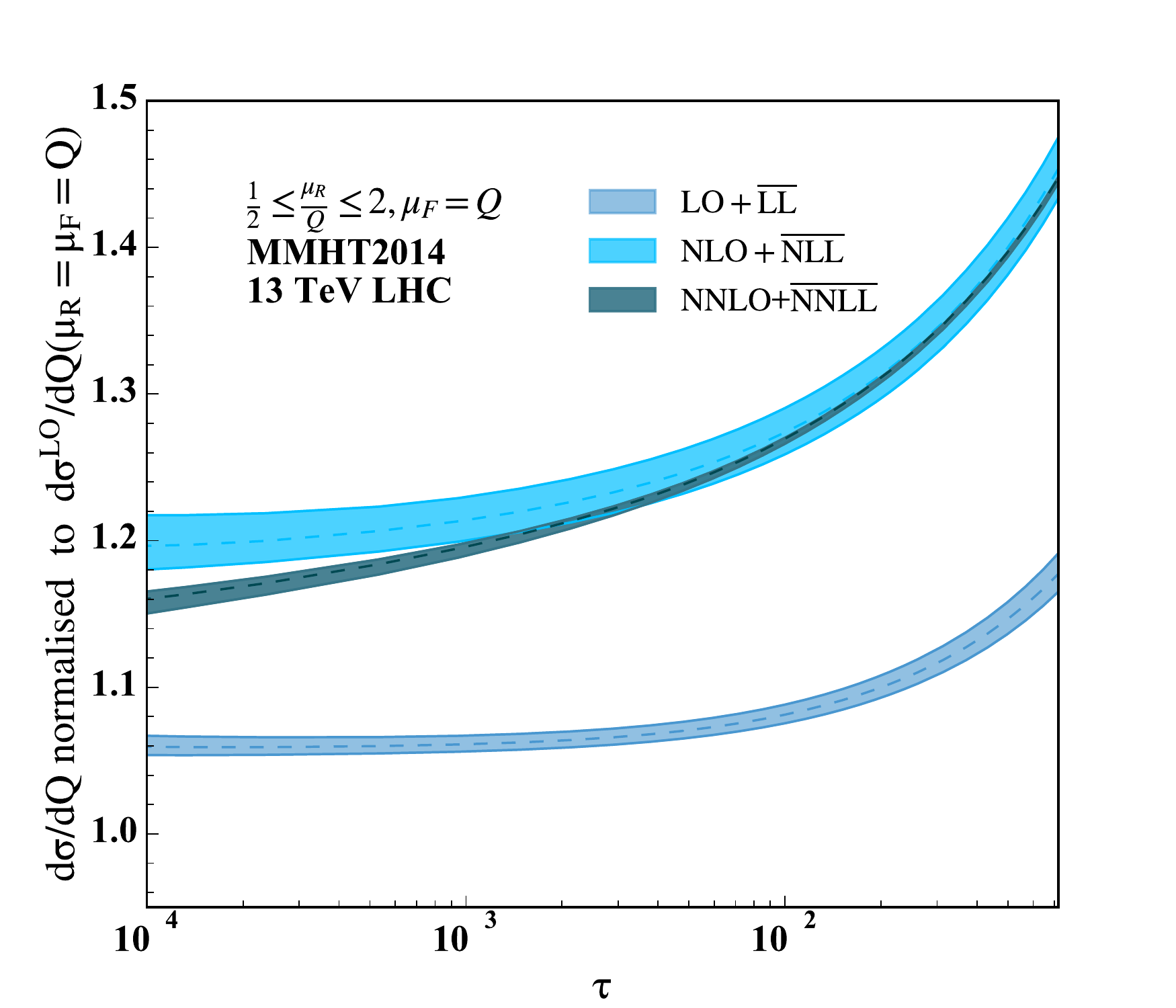}
\caption{$\mu_F$ scale variation (left) of the resummed results  with $\mu_R$ held fixed and $\mu_R$ scale variation (right)  with $\mu_F$ held fixed at $Q$ for 13 TeV LHC }
\label{fig:muF}
\end{figure}

In order to have a better understanding, we study the $\mu_F$ and $\mu_R$ scale variations separately as a function of $\tau$ in Fig. \ref{fig:muF}. The error band due to $\mu_F$ variation has close resemblance to those of 7-point scales, suggesting that the uncertainty is largely due to $\mu_F$ variations. This is sensible, since the $\mu_F$-scales compensate between different partonic channels, which is not possible in this case due to the lack of off-diagonal resummed results. This is further emphasised by the $\mu_R$-variation plot, where the partonic channels do not mix. We see the predictions are less sensitive to $\mu_R$ scale as expected. This essentially hints towards the importance of off-diagonal resummation, which requires further study.

\bibliographystyle{apsrev}
\bibliography{bibliography}

\end{document}